\documentclass[runningheads]{llncs}

\usepackage{url}
\usepackage{graphicx}

\usepackage{multirow}
\usepackage[colorinlistoftodos, textwidth=2.0cm]{todonotes}

\usepackage{xspace}
\newcommand{\mathsymbol}[2]{\newcommand{#1}{\ensuremath{\mathit{#2}}\xspace}}

\mathsymbol{\FOL}{\mathcal{L}}
\mathsymbol{\T}{\mathcal{T}}
\newcommand{\tool}{{\sc MLS-Plan}\xspace}
\newcommand{\htntool}{{\sc HTN-SPlan}\xspace}
\newcommand{\automl}{Auto-ML\xspace}
\newcommand{\autoweka}{\textit{Auto-WEKA}}
\newcommand{\tpot}{\textit{TPOT}}

\newcommand{\autoscikit}{\textit{auto-sklearn}}
\newcommand{\autosklearn}{\autoscikit}

\begin{document}
%
% paper title
% can use linebreaks \\ within to get better formatting as desired
\title{Automated Machine Learning Service Composition}

% author names and affiliations
% use a multiple column layout for up to two different
% affiliations

\author{Felix Mohr, Marcel Wever, Eyke H\"ullermeier}
%\IEEEauthorblockN{Felix Mohr, Marcel Wever, Eyke H\"ullermeier}
%\IEEEauthorblockA{Heinz Nixdorf Institute\\
%Paderborn University\\
%Paderborn, Germany\\
%$\langle$firstname$\rangle$.$\langle$lastname$\rangle$@upb.de}
%\and

\institute{Paderborn University}

% use for special paper notices
%\IEEEspecialpapernotice{(Invited Paper)}

% make the title area
\maketitle

\begin{abstract}
Automated service composition as the process of creating new software in an automated fashion has been studied in many different ways over the last decade.
However, the impact of automated service composition has been rather small as its utility in real-world applications has not been demonstrated so far.
This paper presents \tool, an algorithm for automated service composition applied to the area of machine learning.
Empirically, we show that \tool is competitive and sometimes beats algorithms that solve the same task but not benefit of the advantages of a service model.
Thereby, we present a real-world example that demonstrates the utility of automated service composition in contrast to non-service oriented solutions in the same area.

\end{abstract}

% For peer review papers, you can put extra information on the cover
% page as needed:
% \ifCLASSOPTIONpeerreview
% \begin{center} \bfseries EDICS Category: 3-BBND \end{center}
% \fi
%
% For peerreview papers, this IEEEtran command inserts a page break and
% creates the second title. It will be ignored for other modes.

\section{Introduction}
Automated service composition as the process of creating new software in an automated fashion has been studied in many different ways over the last decade \cite{mohr2016}.
	The most commonly addressed problem is the composition or configuration of a single process either by instantiating or refining an abstract workflow \cite{berardi2003,zeng2003,DBLP:conf/semweb/WuPSHN03} or by creating such a process from scratch given some behavior description in terms of preconditions and effects \cite{klusch2005,hoffmann2009message,mohr2015scc}.
	
In the last years, much of the euphoria about automated composition has disappeared.
	First, services in the real world did not appear so nicely described as expected, which rules out many approaches relying on such descriptions.
	Even though much functionality is available as services, semantic descriptions, e.g., in OWL-S, are rare.
	Second, even for approaches not relying on such assumptions, automated service composition has been resolved mostly on toy examples and not shown to be relevant in real world scenarios.
	In fact, there is only a hand full of approaches that leave the \emph{description} level at all to work with actually implemented services \cite{mcilraith2002,waldinger2001,narayanan2002}.
	However, even these are rather artificial and not real-\emph{world} services.
%We are not aware of any fully automated composition approach that creates executable services.
    
The main contribution of this paper is to demonstrate that automated service composition can outperform manual or automated non-service-oriented software composition in real world applications.
	The domain of the considered service composition problem is \emph{automated machine learning}.
	More precisely, given some sample data, the task is to compose a machine learning pipeline (consisting of machine learning services) that maximizes the classification accuracy over new data from the same source.
	Up to now, this problem has only been tackled by framework-specific tools such as \autoweka{} \cite{DBLP:conf/kdd/ThorntonHHL13,DBLP:journals/jmlr/KotthoffTHHL17} and \autoscikit{}\cite{DBLP:conf/nips/FeurerKESBH15}.
	However, virtually all of the algorithms are available not only in the libraries of those frameworks but also as services offered by commercial platforms\footnote{For the reviewer: algorithmia.com is such a provider, but we do not want to advertise specific providers, so this will not be contained in the paper}.
% * <wever@mail.upb.de> 2018-03-07T20:38:18.043Z:
% 
% Link/Referenz zu diesen Plattformen?
% 
% ^.

This paper augments our approach sketched in \cite{mohr2018} by a detailed technical description and an empricial evaluation that shows the benefits of using services for the purpose of automated machine learning.    
Similar to existing composition approaches \cite{DBLP:conf/semweb/WuPSHN03,sohrabi2006web,sohrabi2009web}, our approach builds on top of hierarchical planning.
	The main difference to existing approaches is that the search process is based on performance measures that are acquired from the execution of composition candidates.
	Since classical planners do not support such a guidance, we implemented a new planner, \tool, which is available for public\footnote{\url{https://github.com/fmohr/ML-Plan}}.
% * <wever@mail.upb.de> 2018-03-07T20:39:23.449Z:
% 
% Hier: ML-Plan oder MLS-Plan also \tool ?
% 
% ^.
We support our claim with an empirical evaluation in which we compare our approach against traditional tools and a non-service-oriented version of our own tool.

\section{Background and Motivation}
Automated service composition is often reduced to AI planning \cite{mohr2016}, and hierarchical automated service composition conducts such a reduction to hierarchical planning \cite{DBLP:books/daglib/0014222}.
% * <wever@mail.upb.de> 2018-03-07T20:39:59.533Z:
% 
% Cite fehlt
% 
% ^.
The core idea of hierarchical task planning (HTN) is to iteratively break down an initially given complex task into new sub-tasks, which may also be complex or simple (no need of further refinement).
The complex tasks are recursively decomposed until only simple tasks remain.
This is comparable, for example, to deriving a sentence from a context-free grammar, where complex tasks correspond to non-terminals and simple tasks are terminal symbols.

There have been several approaches to hierarchical automated service composition.
All these approaches are based on the composite process model in OWL-S.
Roughly speaking, a composite process is just an abstract process consisting of a control flow that contains invocations of other service operations.
The service composition problem is represented by a description of such a composite process, which is equivalent to the initial complex task of the HTN problem with only one possible refinement corresponding to the service process.
Existing services are either also composite processes, which are translated to complex tasks, or atomic service, which are translated to simple tasks.
The very initial works on this topic \cite{sirin2004} were concerned with deriving \emph{any} valid composition, which is usually a trivial undertaking.
Paik et al. slightly extended this setup by considering additional logic constraints on plans \cite{paik2007automatic}.
Follow-up work by Sohrabi et al. considered the fact that clients may order plans based on preferences that can be expressed in terms of logic conditions achieved by a process \cite{sohrabi2006web,sohrabi2009web,DBLP:journals/re/LiaskosMSM11}.

Curiously, there has been almost no work on optimizing the composition quality in terms of Quality of Service (QoS).
While QoS optimization has been studied a lot in automated service composition in general, there is almost no such work on hierarchical composition approaches.
Indeed, the preferences over plans in the work of Sohrabi are also induced by an implicit quality.
However, the only work that optimize QoS typically considered in automated service composition such as runtime, cost, etc. has been presented by \cite{chen2009markov}.

In this paper, we consider the problem of finding a composition with best (numeric) quality with the limitation that this quality measure can \emph{not} be aggregated from the services contained in the composition.
So our setup is similar to the one of \cite{chen2009markov} in that we assume a numeric quality to be optimized, but the difference is that there is no (known) way to statically assign qualities to existing services and compute the quality of a composition from these qualities.
Instead, the quality of a composition can be accessed by a benchmark that executes the composition and observes the desired quality.

A highly relevant real world example for such a problem is automated machine learning pipeline construction.
In a nutshell, the classification problem in machine learning is to learn a (non-deterministic) relation between \emph{instances} $\mathcal{X}$ and \emph{class labels} $\mathcal{Y}$.
Instances are described in terms of numeric or categorical attributes called features, and a set of such instances is given, each of which is additionally associated with a class label.
% * <wever@mail.upb.de> 2018-03-07T20:45:31.397Z:
% 
% Die echte Ground Truth kennt man u.U. doch gar nicht => Noise
% 
% ^.
The goal is to establish a new function $h: \mathcal{X} \rightarrow \mathcal{Y}$, called the hypothesis, such that $h$ maximizes the ratio of correctly predicted class labels for \emph{new} instances.
% * <wever@mail.upb.de> 2018-03-07T20:46:25.072Z:
% 
% Hier wird erst der Out-Of-Sample Error beschrieben, der jedoch nicht berechnet werden kann, dann Anforderung an die measure dass sie berechenbar sein muss.
% 
% ^.
The hypothesis may be a single classification algorithm such as a decision tree, a neural network, or a support vector machine, or it may be a whole pipeline consisting of (possibly complex) pre-processing algorithms followed by such a classifier.
Finding such a pipeline automatically is a hot topic in machine learning, but all existing solutions focus on platform-dependent frameworks \cite{DBLP:conf/kdd/ThorntonHHL13,DBLP:conf/nips/FeurerKESBH15}.
However, service implementations whose communication is based on HTTP exist for all of these algorithms, so it is also possible to solve the \automl{} problem as a service composition problem.

Note that the true error rate cannot even be computed exactly but only estimated.
The true error of a pipeline is the average error produced over all data points that \emph{exist}, but only a finite sample of such points is available.
Hence, one estimates the out-of-sample error by keeping a validation set of the initial data and using it for estimating that error.

We claim that solving the Auto-ML problem as a service composition can be significantly better than sticking to a single algorithm framework such as WEKA or scikit-learn.
The main reason for this is that the portfolio of implementations from which one can select algorithms is much broader.
WEKA implements the algorithms in Java, and scikit-learn implements them in Python.
Without the encapsulation into services, it is not easily possible to use algorithms of both during optimization.

\section{Hierarchical Service Composition Approach}\label{sec:approach}
\begin{figure}[t]
    \centering
	\includegraphics[width=.6\columnwidth]{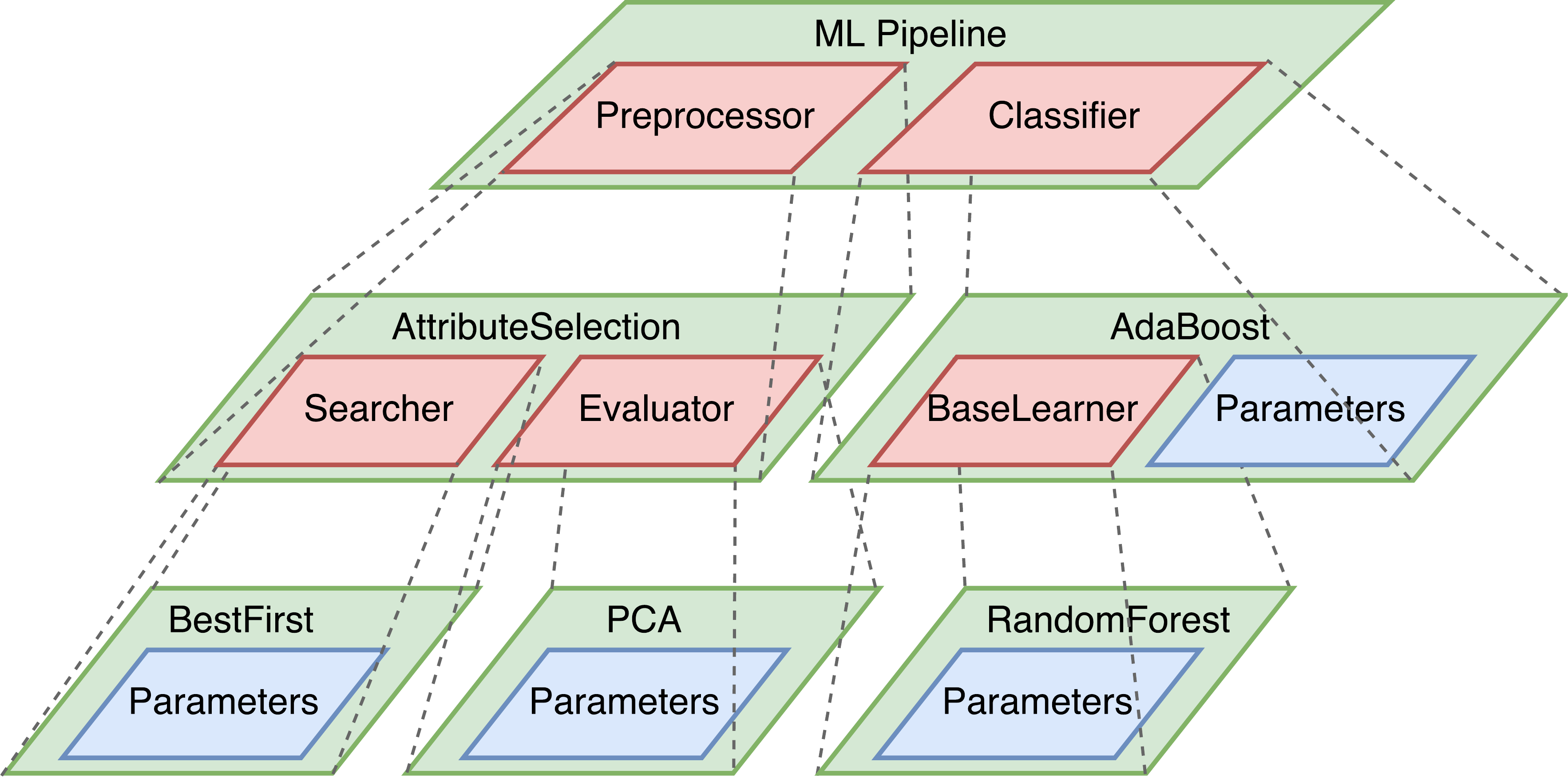}
    \label{fig:decomposition}
    \caption{Visualization of the hierarchical structure of an machine learning pipeline}
\end{figure}

In this section, we describe the hierarchical planning formalism we use to create the \emph{constructor} of the composed service.
Like most other approaches, our composition algorithm does not directly compose an entire service but a \emph{process}.
We assume that the target service has already been defined in the form of a template with one missing process (the constructor).
For example, we already defined how the machine learning pipeline service will work but not the atomic services on which it will rely.
The solution of the composition problem derived with the approach in this section will be injected into that template in order to obtain a ready-to-use service; Fig. \ref{fig:decomposition} sketched this hierarchical template structure.
This section explains how this construction process is created, and the following section will explain how the executable service is obtained from it.

\subsection{Formalization of HTN Planning}

%We adopt the syntax and semantic of early HTN approaches to service composition \cite{sirin2004}.

As for any planning formalism, our basis is a logic language \FOL and planning operators defined in terms of \FOL.
The language \FOL has first-order logic capacities, i.e., it defines an infinite set of variable names, constant names, predicate names, function names and quantifiers and connectors to build formulas.
A \emph{state} is a set of ground literals; i.e., it does not contain unquantified variable symbols.
We do \emph{not} adopt the closed-world assumption.

Constants, functions, and predicates of \FOL may stem from a \emph{theory}.
A theory \T defines constants, functions, and predicates and how these are to be interpreted.
Predicates not contained in \T behave like normal predicates in classical planning.
That is, \FOL consists of the elements of \T together with uninterpreted predicates and constants.
In the formalism, we use \T as a formula itself.

An \emph{operator} is a tuple $\langle \mathit{name_o, I_o, O_o, P_o, E_o^+, E_o^-}\rangle$ where $name_o$ is a name, $I_o$ and $O_o$ are parameter names described inputs and outputs, $P_o$ is a formula from \FOL constituting its preconditions and $E_o^+$ and $E_o^-$ are sets of conditional statements $\alpha \rightarrow \beta$ where $\alpha$ is a formula over \FOL conditioning the actual effect $\beta$, which is a set of literals from \FOL to be added or removed.
Free variables in $P_o$ must be in $I_o$ and free variables in $E^+_o$ and $E^-_o$ must be in $I_o \cup O_o$.

The semantics of the planning domain are as follows.
An \emph{action} is an operator whose input and output variables have been replaced by constants; we denote $\mathit{P_a, E^+_a}$, and $E^-_a$ as the respectively replaced preconditions and effects.
An action $a$ is \emph{applicable} in a state $s$ under theory \T iff $s,\T \models P_a$ and if none of the output parameters of $a$ is contained in $s$.
Applying action $a$ to state $s$ changes the state in that, for all $\alpha \rightarrow \beta \in E^+_a$, $\beta$ is added to $s$ if $s,\T \models \alpha$; analogously, $\beta$ is removed if such a rule is contained in $E^-_a$.
A \emph{plan} for state $s_0$ is a list of actions $\langle a_0,..,a_n\rangle$ where $a_i$ is applicable and applied to $s_i$; here, $s_{i+1}$ is obtained by applying $a_i$ to $s_i$.

On top of this formalism, we build a hierarchical model \cite{DBLP:conf/aips/2016}.
A task network is a partially ordered set $T$ of tasks.
A task $t(v_0, .., v_n)$ is a name with a list of parameters, which are variables or constants from \FOL.
For example, $setPreprocessor(pl)$ could be the task of choosing and setting the preprocessing algorithm for the pipeline object pl.
A task named after an operator is called \emph{primitive}, and \emph{complex} else. A task whose parameters are constants is ground.

While primitive tasks are realized canonically by an operation, complex tasks need to be decomposed by \emph{methods}.
A method $m = \langle name_m, t_m, I_m, O_m, P_m, T_m \rangle$ consists of its name, the (non-primitive) task $t_m$ it refines, the input and output parameters $I_m$ and $O_m$, a logic formula $P_m \in \mathcal{L}$ that constitutes the method's precondition, and a task network $T_m$ that realizes the decomposition.
The preconditions may, just as in the case of operations, contain interpreted predicates and functional symbols from the theory \T.

A method instantiation $m$ is a method where inputs and outputs have been replaced by planning constants.
$m$ is \emph{applicable} in a state $s$ under theory \T iff $s,\T \models P_m$ and if none of the output parameters of $m$ is contained in $s$.

An HTN planning problem is a tuple $P = \langle O, M, s_0, N\rangle$ where $O$ is a set of operations as above, $M$ is a set of methods, $s_0$ is the initial state, and $N$ is a task network.
%The conditions for a plan $\pi = \langle a_1,..,a_n\rangle$ that is applicable in $s_0$ being a \emph{solution} to $P$ are inductive and based on three cases:
%\begin{enumerate}
	%\item $N$ is empty. $\pi$ is a solution if it is empty
	%\item $N$ has a primitive task $t$ without predecessor in $N$. $\pi$ is a solution if $a_1$ %realizes $t$ and is applicable in $s_0$ and if $\langle a_2,..,a_n\rangle$ is a solution to %$\langle O,M,\tau(s_0,a_1), N \setminus \{t\}\rangle$.
	%\item $N$ has a complex task $t$ without predecessor in $N$. $\pi$ is a solution if there is an %instantiation $\widehat{m}$ of a method $m \in M$ that is applicable in $s_0$ yielding a refined %network $N'$, and $\pi$ is a solution to $\langle O,M,s_0,N'\rangle$.
%\end{enumerate}
%Note that these cases are not mutually exclusive unless $N$ is totally ordered.
An HTN \emph{optimization} problem is an HTN planning problem together with an objective function.
Formally, $P^*$ is an HTN optimization problem iff it is a tuple $P^* = \langle O, M, s_0, N, \phi\rangle$  where $P := \langle O, M, s_0, N\rangle$ is an HTN planning problem and $\phi$ is a real-valued function that assigns a score to any solution of $P$.
A plan $\pi^*$ is a solution to $P^*$ if it is a solution to $P$ and there is no other plan $\hat{\pi}^*$ such that $\phi(\hat{\pi}^*) > \phi(\pi^*)$.

\subsection{The Service Composition Problem}
We assume that the composition domain is described in terms of available services and macros that encode abstract processes.
Services are described by a name and a set of offered operations.
That is, the services are a set $\{s_1,..,s_n\}$ where each service $s_i$ is a tuple $\langle name_i, \{op^i_1,..,op^i_{m_i}\}\rangle$, and each operation $op^i_j$ is described like the planning operators of the HTN problem by a name, inputs, outputs, preconditions, and effects.
Macros are generic process templates describing reasonable process abstractions in the domain.
Every macro consists of a name, conditions under which it can be applied, and its actual process.
 The elements of the process are calls to service operations or other macros, i.e., names of service operations or macros together with bindings for the data objects used in the inputs and outputs.
In this paper, we only consider sequential macros, i.e., sequential processes, but note that if-else-statements can be easily encoded by having a separate macro for each case. 
 
Operations can be called either directly on a service or on a service \emph{instance}.
As in object oriented programming, we assume a class-instance-model; every service constitutes the class of all its instantiations.
For example, a neural network service is the class for all its concrete instantiations (each of which will represent a different network).
Services may have a constructor that creates a new instance and returns the resource for that instance to the invoker.
The service operation calls in the macros are then either calls to static operations of a service (operations that do not depend on a particular state) or calls to the operations of a service \emph{instance}.

Compositions are sequences of service operation invocations.
That is, a composition is a sequence of pairs $(o_i,b_i)$ where $o_i$ is a service (instance) operation, and $b_i$ is a function that maps each input of $o_i$ to outputs of earlier invocations or inputs of the overall process.
Like in other approaches, non-sequential compositions are not considered\footnote{The SHOP2-encoding of Sirin et al. allows for non-sequential composite processes during the composition, but the eventually returned composition is also sequential}.

A service query consists of three parts.
First, it contains a task network as described above.
Second, it defines initial information about the objects that will be passed to the network.
Tasks in that network correspond either to calls to service operations (primitive tasks), or to calls to a macro, which needs to be configured (complex tasks).
Third, it defines an objective function that assigns a score to each possible solution candidate.
This function is not given in a closed-form representation but as a reference to an invocable routine.

The considered \emph{service composition problem} is then described by a triplet of services, macros, and a service query.
Intuitively, a solution candidate to this problem is a composition obtained by recursive replacements of macros by operation calls or other macros such that the precondition of each operation is satisfied in the moment of execution.
A composition is an optimal solution if it is a solution candidate and if no other solution candidate receives a higher score from the objective function.

\subsection{Translating the Service Composition Problem to HTN}
The translation of a service composition problem to an HTN planning problem is analogous to the one in \cite{sirin2004} except two modifications.
First, the fact that we distinguish services from service instances requires a small modification.
There is a clear correspondence between macros and methods on one hand, and service operations and HTN operators on the other hand, so the translation seems to be canonical.
However, allowing to create new service instances (during planning) also means to allow that new operations are created during planning.
It is not immediately clear how one should treat this situation.
Second, we also need to translate the objective function, which does not exist in previous approaches.

The first point can be handled with a simple trick in that we treat all operations as if they were static and add the service instance reference as an additional input.
The different ``versions'' of an operation $o$ of a service $s$ for different instances of $s$ are not really different in their functionality but just deviate in the service instance on which they are invoked.
Hence, instead of adding new operations we just assume that instance-specific operations have an additional and distinguished input that represents a handle for the service instance on which it should be invoked.
Of course, the handle used for a service instance is exactly the reference returned from the constructor of the respective service.

Given the correspondence between service operations and HTN operators, translating the objective function comes down to a simple wrapper.
The only thing this wrapper has to do is to map planning action syntax to service operation invocation syntax.
If this functionality is available, a solution of the HTN problem can be converted into a composition, which can then be executed by the objective function.

Note that we trade the assumption that no service is both information-gathering and world-altering\cite{sirin2004} by the assumption that the execution of world-altering services does not affect the execution of other services or compositions.
Sirin et al. make the first of these two assumptions, because they want to execute \emph{some} (the information-gathering) services during planning.
In order to avoid side-effects during planning, they forbid that such services alter the world.
However, our setup precisely requires to execute entire compositions, so this assumption is not longer needed and reliefs us from the tedious distinction of knowledge effects and physical effects introduced in OWL-S.
On the contrary, we need the assumption that the compositions do not alter the world in such a way that the execution of other compositions is affected.

\subsection{The Planning Algorithm}
\begin{figure}[t]
    \centering
	\includegraphics[width=0.5\columnwidth]{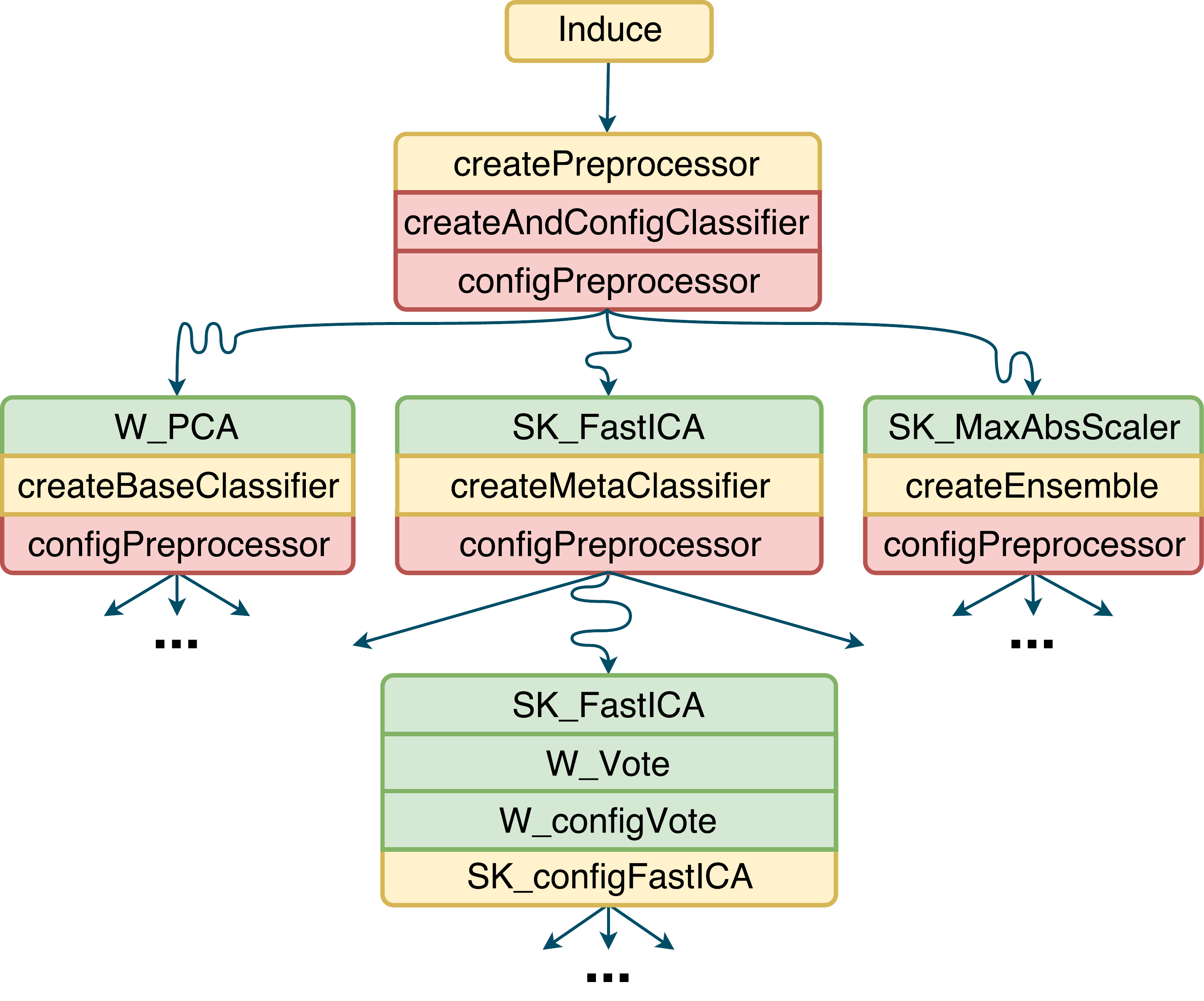}
    \caption{Excerpt of the search graph of the HTN planner}
    \label{fig:task-network}
\end{figure}

Like all planning problems, HTN problems are solved using graph search algorithms.
The (hierarchical) planning problem induces a (possibly infinite) search graph, which is represented by a distinguished root node, a successor generator function, and a goal-test function.
The successor generator creates the successor nodes for any node of the graph, and the goal-test decides whether a node is a goal.
Most HTN planners perform a \emph{forward-decomposition}, which means that they create one successor for each possible decomposition of the first unsolved task in the list of remaining tasks.
In every child node, the list of remaining tasks is the previous list of tasks where the decomposed task is replaced by the list that represents the respective decomposition.
The resulting search graph is sketched in Fig. \ref{fig:task-network} where every box shows a list of tasks (green ones are simple, the yellow one is the next complex task to be decomposed, and the red ones are complex to be resolved later).
A node is a goal node if all remaining tasks are simple.
A standard graph search algorithm can then be used to find a path from the root to a goal node.

%Unfortunately, the compiled HTN optimization problem cannot be reasonably approached using standard planners such as SHOP2 \cite{DBLP:journals/corr/abs-1106-4869}.
%The problem is that standard planners, which usually apply A* search, assume that the cost of a solution (e.g., expected prediction accuracy of a classifier) can be computed from the descriptions of the plan elements.
%However, this precisely contradicts our assumption that such an aggregation is \emph{not} possible.
%One can still apply these tools, but they cannot be informed about the quality of the search graph nodes they traverse; this effectively yields a blind search.

To overcome the limitation of standard solvers to additive cost measures, we developed \htntool, an HTN planner based on arbitrary node evaluation functions.
\htntool realizes a best-first search where each node is labeled with elements of an ordered set (usually real values or vectors with tie breaker).
\htntool makes no assumption (like monotonicity) about the node evaluations or how they are acquired.
Instead, it simply requires that the node evaluation function is provided by the user.
It is then possible to conduct complex computations in order to obtain node evaluations, a property that is missing in classical planners.

Besides classical node evaluation functions, \htntool offers another default node evaluation function based on random path completion as also used in Monte Carlo Tree Search \cite{DBLP:journals/tciaig/BrownePWLCRTPSC12}.
To obtain the evaluation of a node, this strategy draws a fixed number of path completions and evaluates the plan using a given plan evaluation function.
The score assigned to the node is the best score that was observed over these validations in order to estimate the best solution that can be obtained when following paths under the node.

Using this evaluation function, \htntool is an appropriate tool to solve the service composition problem.
The plan evaluation function is the wrapper of the objective function.

It is important to be aware that, in contrast to classical heuristic approaches, \tool does not give any guarantees about the optimality of returned solutions.
This is precisely because it does make no assumptions about the node evaluation function, so there is actually no guarantee possible.
So, strictly speaking, the algorithm does not even return solutions in the narrower sense at all but only solution candidates, because optimality is a solution criterion.

However, this is not a particular weakness of \htntool since, without further assumptions, it is not even \emph{possible} to prove the optimality of a solution without enumerating all candidates.
Unless all solutions have been observed, \emph{every} algorithm is prone to miss the true optimum.

While it is probably possible to make \emph{some} assertions about optimality (usually in the form of bounds), we do not provide such proofs here.
In fact, for the concrete evaluation function based on random completions, some guarantees appear provable since the algorithm is similar to UCT for which bounds have already been shown.
% * <wever@mail.upb.de> 2018-03-07T22:28:53.824Z:
% 
% Cite leer, Referenz fehlt
% 
% ^.
However, proofs for such bounds are way beyond the scope of this work.
We rather focus on experimental analysis and will show that the solutions produced by \tool, even though being usually sub-optimal, still significantly outperform any other solution produced by other algorithms.
% * <wever@mail.upb.de> 2018-03-07T22:29:28.553Z:
% 
% Hier wieder outperform. Und die Ankündigung wiederholt sich hier bereits zum 3. oder 4. Mal
% 
% ^.

\section{Case Study: Machine Learning Pipelines}
In our case study, we consider the domain of automated machine learning (\automl).
\automl aims at automatically selecting and configuring the algorithms of a so-called machine learning pipeline.
Usually, such a pipeline consists of one or more preprocessor algorithms (principal components, imputation, etc.) and a classification algorithm (decision trees, logistic regression, etc.).
State-of-the-art approaches \autosklearn{} and \autoweka{} reduce the combined algorithm selection and hyperparameter optimization  problem to a mere hyperparameter optimization problem, considering the selection of an algorithm for feature preprocessing and a classifier model as additional parameters for a hyperparameter optimization tool. Moreover, these tools are committed to a certain library (e.g., scikit-learn or WEKA) as well as to a specific programming language (e.g. Python or Java). However, these libraries are neither equal nor does one subsume the other. Moreover, implementations of certain machine learning algorithms usually differ significantly so that even for a particular algorithm there might be differences in terms of non-functional requirements, e.g., runtime or even predictive accuracy.

In the following, we describe \tool, the application of \htntool to the \automl problem.
We first describe how we created services out of the existing algorithms and how the execution of compositions works.
We then present an experimental evaluation, which shows that the service-based approach combing WEKA and scikit-learn services is often better than using the same search technique with algorithms of just one library, and it is even mostly competitive with expert approaches such as \autoweka{} and \autoscikit.

%A more natural way of modeling this problem is to interpret each of the algorithms as a service which may be part of a ML pipeline. The latter may incorporate several of those services to accomplish the desired machine learning task. This, is precisely what is done in \tool, enabling the use of algorithm implementations independent of the platform, language, and library. However, as a prerequisite, the implementations of libraries need to be wrapped into services, first, which is described in Section~\ref{ssec:servicification}.

\subsection{Servicification of Existing Libraries}\label{ssec:servicification}
The idea of what we call servicification is to make existing software accessible as a service.
Our contribution is not about this process in general, so we only describe how we convert the learning algorithms relevant for the case study into services.

\begin{figure}[t]
\centering
\includegraphics[width=0.6\columnwidth]{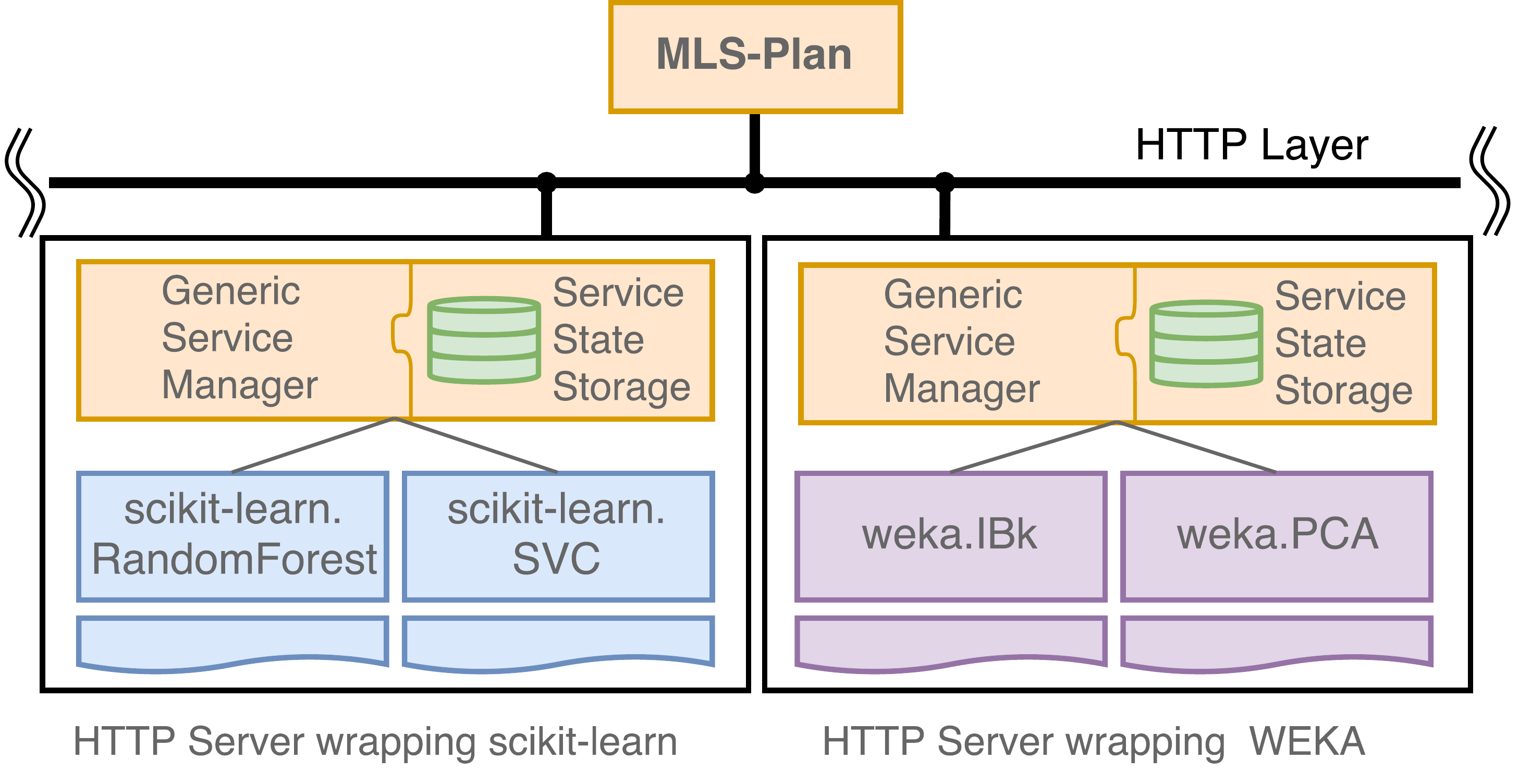}
\caption{Communication via HTTP }
\label{fig:http-layer}
\end{figure}

We enable servicification by so called \textit{Generic Service Managers} (GSM), which are web servers that route HTTP requests to invocations of functions in an object-oriented programming language.
Every algorithm in the considered machine learning frameworks is encoded in its own (Java or Python) class file, so we only consider these two languages here.
A GSM processes requests of the form  \texttt{http://host/classname/method} with which the client can trigger the invocation of a given method of a (generally arbitrary) class.
The parameters are transmitted by GET or POST; in our case, we only have POST requests.
The GSM is generic as it does not have to be tied to a specific library and may create objects via reflection in Java or via \texttt{importlib} and \texttt{getattr} in Python.
For simplicity, we use one GSM for Java classes and one for Python classes (Figure~\ref{fig:http-layer}).
In other words, the set of available services is the set of classes enabled in the GSM, and the service operations correspond to the (enabled) methods of that class.

In our framework, services are generally stateful.
We are aware that there are paradigms that suggest that not only the communication between services but also the services themselves should be stateless.
However, we argue that many services, including those related to machine learning, are more reasonably realized in a stateful manner.
For example, machine learning services should save the model they learned locally instead of exchanging it with the client in order to keep themselves stateless.
Not only are such models sometimes very large, but the model can also be seen as a part of the implementation of the service and, hence, the service provider might not want to share the model with the client.

Stateful services are realized by a distinction between services and their instances.
This is analogue to classes and objects in programming languages: services correspond to classes, and service instances to objects of those classes.
A service is then instantiated by simply creating an object of the respective class.
The GSM interprets requests of the form \texttt{http://host/classname/new} as commands to create new service instances.
The response to such requests is the URL by which the new service \emph{instance} is reachable, which, by  convention, is the same URL as for the service itself plus some ID.
Operations of a service instance can then be accessed via \texttt{http://host/classname/id/method}.

To make the service instances persistent, the GSM serializes the respective objects to its disk.
When the Java or Python objects are created on an invocation of \texttt{http://host/classname/new}, their lifetime is limited by the lifetime of the web server worker thread.
Hence, the GSM serializes the created objects on the server disk in order to have an unlimited lifetime; this storage is called the Service State Storage (S3).
Of course, this requires that the class to be servicified actually is serializable.
However, all of the considered algorithm classes \emph{are} serializable, so this is no limitation in the given domain.

%Furthermore, the GSM is given access to some storage system, which we refer to as \textit{Service State Storage} (3S), enabling the GSM to store serialized objects. In order to uniquely identify different service instances, each service instance is assigned a reference ID token.

\subsection{Turning Composition Problem Solutions into Services}
The solutions to the composition problem defined in Section \ref{sec:approach} are \emph{processes} but not services.
For example, we can compose a process that describes how a machine learning pipeline is configured, but the process does not correspond to such a pipeline itself.

To obtain composed services from solutions of the composition problem, we use service \emph{templates}.
The template contains the code for the different \emph{operations} of the target service \emph{except} its constructor, which will be used to configure the (basic) services on which the (composed) service will rely.
For example, the template may contain an operation to \emph{train} the machine learning pipeline and to \emph{predict} the label of new instances based on preprocessing and classification services $s1$ and $s2$ respectively where $s1$ and $s2$ are not defined in the code but expected to be defined in the (yet empty) constructor.
The solution of the composition problem is translated into the \emph{constructor} of that template in order to obtain an executable program.
Using the GSM, the composed service can be accessed as a service in turn.

In the following, we refer to \emph{compositions} as the processes of \emph{any} operation of the composed service.
For example, in the machine learning case, there is one composition for the constructor, one for training the pipeline, and one for predicting new labels.
Syntactically, the processes of the training and prediction compositions are fixed and do not depend on the solution of the composition problem.
However, since these operations will rely on external basic services configured in the constructor their \emph{behavior does} depend on the concrete composition injected into the constructor.

\subsection{Processing Compositions via Choreography}
An efficient execution of compositions requires that the participating services communicate via a choreography protocol.
In our case study, composed processes include, for example, the application of a preprocessor followed by the training of the actually used classification algorithm.
The data passed to these services can easily reach several hundreds MB or even some GB, which makes a zig-zag-communication with the client unacceptably slow.

To this end, every service operation receives, in addition to its usual arguments, the entire composition.
The service then only sends its result back to the client if it is the last operation in the composition; otherwise it directly sends its data to the next service operation in the choreography.
Our current implementation only supports sequential compositions where the input of one service is provided by the preceding service without hops in the data-flow, which is sufficient for the case study and keeps the implementation simple.
% * <wever@mail.upb.de> 2018-03-07T15:49:18.456Z:
% 
% Was bedeutet "without hops in the data-flow"?
% 
% ^ <wever@mail.upb.de> 2018-03-07T15:55:38.715Z.

The execution logic for compositions is also contained in the GSMs.
That is, GSMs can not only process service invocations but also entire compositions.
To this end, they receive a single service operation call together with a composition.
They identify the position of the invoked service within the composition, execute it and send the result either to the client or to the subsequent service if one exists.

\subsection{Experimental Setup}\label{ssec:experiment-setup}
\newcommand{\numdatasets}{21}
\newcommand{\numruns}{20}
\newcommand{\numcandidates}{6}

To asses the question of what the benefit of using services is, we evaluate \tool{} incorporating both scikit-learn (Python) and WEKA (Java) in the form of HTTP services, comparing it to itself limited to use only scikit-learn or weka respectively. Additionally, we compare \tool{} to other \automl{} tools such as \autoweka{} (WEKA, Java), \autosklearn{}, and \tpot (scikit-learn, Python) \cite{OlsonGECCO2016} to put our results into context of state-of-the art solutions.
%the state-of-the-art \automl{} tools \autoweka{} (weka, Java) and \autoscikit{} (scikit-learn, Python). Furthermore, we compare \tool{} to two other \automl{} tools T-POT (scikit-learn, Python) and {\sc RecipeML} (scikit-learn, Python) and lastly to the generic planning tool JSHOP2 (weka, Java).

\begin{table*}[ht]
\bgroup
\def\arraystretch{1.0} % row height: 1 is the default
\setlength\tabcolsep{4pt} % col width: 6 is the default
\resizebox{\textwidth}{!}{
\begin{tabular}{l|r|r|r||r|r|r||r|r|r}
\hline
Dataset &\#Inst. &\#Att.  &\#Classes& \multicolumn{1}{c|}{\tool} & \multicolumn{1}{c|}{\tool\ (W)}& \multicolumn{1}{c||}{\tool\ (S)}& \multicolumn{1}{c|}{\autoweka}& \multicolumn{1}{c|}{\tpot} & \multicolumn{1}{c}{\autoscikit}\\\hline
% MLS-Plan => ML-Plan(W) => Auto-WEKA => auto-sklearn
{\sc abalone} & 4177 & 8 & 28 & \textbf{73.45$\pm$1.2}   & 73.65$\pm$1.3 $\phantom{\circ}$  & 78.02$\pm$1.1 $\bullet$  & 73.46$\pm$1.1 $\circ$  & 73.12$\pm$1.0 $\circ$  & 80.09$\pm$6.9 $\bullet$ \\
{\sc amazon} & 1500 & 10000 & 50 & \textbf{26.28$\pm$13.7}   & 38.23$\pm$14.1 $\bullet$  & 98.0$\pm$0.0 $\bullet$  & 51.72$\pm$2.7 $\bullet$  & -\phantom{$\bullet$} & 28.57$\pm$5.4 $\phantom{\circ}$ \\
{\sc car} & 1728 & 6 & 4 & \textbf{1.28$\pm$1.3}   & 1.82$\pm$1.5 $\bullet$  & 1.78$\pm$2.0 $\bullet$  & 0.66$\pm$0.4 $\circ$  & 0.28$\pm$0.3 $\circ$  & 1.56$\pm$0.7 $\phantom{\bullet}$ \\
{\sc convex} & 58000 & 784 & 2 & 46.5$\pm$1.5   & \textbf{32.29$\pm$12.8} $\circ$  & -\phantom{$\bullet$} & 46.83$\pm$0.4 $\bullet$  & -\phantom{$\bullet$} & 16.12$\pm$0.9 $\circ$ \\
{\sc credit-g} & 1000 & 20 & 2 & \textbf{24.53$\pm$2.35}   & 24.79$\pm$1.3 $\phantom{\circ}$  & 25.4$\pm$2.3 $\phantom{\bullet}$  & 26.5$\pm$2.4 $\bullet$  & 25.17$\pm$4.9 $\phantom{\bullet}$  & 27.29$\pm$0.7 $\bullet$ \\
{\sc dexter} & 600 & 20000 & 2 & \textbf{5.56$\pm$2.4}   & 9.01$\pm$3.5 $\bullet$  & 5.71$\pm$0.6 $\phantom{\bullet}$  & 11.44$\pm$2.8 $\bullet$  & -\phantom{$\bullet$} & -\phantom{$\bullet$}\\
{\sc dorothea} & 1150 & 100000 & 2 & \textbf{6.61$\pm$1.9}   & 9.83$\pm$0.0 $\bullet$  & -\phantom{$\bullet$} & -\phantom{$\bullet$} & -\phantom{$\bullet$} & -\phantom{$\bullet$}\\
{\sc gisette} & 7000 & 5000 & 2 & \textbf{2.63$\pm$0.3}   & 2.99$\pm$0.5 $\bullet$  & -\phantom{$\bullet$} & 3.90$\pm$0.4 $\bullet$  & -\phantom{$\bullet$} & 2.60$\pm$0.4 $\phantom{\bullet}$ \\
{\sc glass} & 214 & 9 & 7 & 25.09$\pm$4.7   & \textbf{23.58$\pm$7.6} $\phantom{\circ}$  & 24.82$\pm$4.7 $\phantom{\circ}$  & 25.02$\pm$6.0 $\phantom{\bullet}$  & 24.13$\pm$5.4 $\circ$  & 26.12$\pm$8.4 $\phantom{\bullet}$ \\
{\sc ionosphere} & 351 & 34 & 2 & 6.16$\pm$2.2   & 7.70$\pm$2.8 $\bullet$  & \textbf{6.16$\pm$2.0} $\phantom{\bullet}$  & 6.86$\pm$3.0 $\phantom{\bullet}$  & 6.37$\pm$1.5 $\phantom{\bullet}$  & 7.30$\pm$2.5 $\bullet$ \\
{\sc iris} & 150 & 4 & 3 & 5.71$\pm$3.1   & \textbf{5.06$\pm$3.3} $\phantom{\bullet}$  & 6.36$\pm$3.1 $\phantom{\bullet}$  & 4.38$\pm$3.7 $\phantom{\bullet}$  & 1.19$\pm$1.2 $\circ$  & 6.16$\pm$4.0 $\bullet$ \\
{\sc letter} & 20000 & 16 & 26 & 3.65$\pm$1.1   & 4.17$\pm$0.3 $\bullet$  & \textbf{3.60$\pm$1.4} $\phantom{\bullet}$  & 5.19$\pm$1.6 $\bullet$  & -\phantom{$\bullet$} & 4.89$\pm$0.4 $\bullet$ \\
{\sc madelon} & 2600 & 500 & 2 & 20.53$\pm$5.9   & \textbf{19.47$\pm$2.7} $\phantom{\bullet}$  & 26.17$\pm$5.4 $\bullet$  & 25.52$\pm$3.9 $\bullet$  & 100.0$\pm$0.0 $\bullet$  & 17.68$\pm$2.1 $\circ$ \\
{\sc page-blocks} & 5473 & 10 & 5 & 2.81$\pm$0.5   & \textbf{2.55$\pm$0.3} $\circ$  & 2.87$\pm$0.3 $\phantom{\bullet}$  & 2.68$\pm$0.3 $\phantom{\circ}$  & 2.49$\pm$0.4 $\phantom{\circ}$  & 2.73$\pm$0.2 $\phantom{\circ}$ \\
{\sc secom} & 1567 & 590 & 2 & \textbf{6.42$\pm$0.0}   & 6.84$\pm$0.2 $\bullet$  & 6.93$\pm$0.3 $\phantom{\bullet}$  & 6.55$\pm$0.4 $\phantom{\circ}$  & 6.42$\pm$0.0 $\phantom{\circ}$  & 6.58$\pm$0.3 $\phantom{\circ}$ \\
{\sc segment} & 2310 & 19 & 7 & \textbf{2.77$\pm$0.7}   & \textbf{2.77$\pm$0.6} $\phantom{\circ}$  & 2.98$\pm$0.8 $\phantom{\bullet}$  & 2.04$\pm$0.5 $\circ$  & 1.99$\pm$0.7 $\circ$  & 2.68$\pm$1.0 $\phantom{\circ}$ \\
{\sc semeion} & 1593 & 256 & 10 & 7.27$\pm$1.3   & \textbf{8.57$\pm$0.7} $\circ$  & 7.51$\pm$1.3 $\phantom{\bullet}$  & 12.59$\pm$4.0 $\bullet$  & 5.69$\pm$1.3 $\circ$  & 6.74$\pm$1.2 $\circ$ \\
{\sc vowel} & 990 & 13 & 11 & \textbf{2.34$\pm$1.1}   & 4.33$\pm$1.6 $\bullet$  & 5.15$\pm$3.1 $\bullet$  & 10.07$\pm$8.2 $\bullet$  & 2.01$\pm$0.6 $\circ$  & 6.83$\pm$2.6 $\bullet$ \\
{\sc waveform} & 5000 & 40 & 3 & \textbf{13.03$\pm$0.6}   & 13.51$\pm$0.7 $\bullet$  & 14.93$\pm$1.7 $\bullet$  & 13.35$\pm$0.8 $\phantom{\circ}$  & 13.08$\pm$0.5 $\phantom{\circ}$  & 13.83$\pm$0.8 $\phantom{\circ}$ \\
{\sc winequality} & 4898 & 11 & 11 & 36.32$\pm$2.81   & \textbf{33.86$\pm$0.9} $\circ$  & 36.97$\pm$1.3 $\phantom{\circ}$  & 33.69$\pm$1.9 $\circ$  & 32.37$\pm$1.1 $\circ$  & 36.89$\pm$1.7 $\phantom{\circ}$ \\
{\sc yeast} & 1484 & 8 & 10 & \textbf{41.61$\pm$2.9}   & 43.46$\pm$4.8 $\bullet$  & 42.48$\pm$3.3 $\phantom{\bullet}$  & 39.72$\pm$2.3 $\circ$  & 38.73$\pm$2.6 $\circ$  & 39.43$\pm$1.7 $\circ$ \\
\hline
    \end{tabular}}
    \caption{Means and standard deviation of 0-1 loss. Each entry represents the mean and standard-deviation over \numruns{} runs with different random seeds. }
    \label{tbl:results}
    \egroup

\end{table*}

Results were obtained by carrying out \numruns{} runs on \numdatasets{} datasets with a timeout of 1h.
All of the used datasets can be found in the OpenML\footnote{https://www.openml.org/} dataset repository.
The significance of an improvement or degradation was determined using the t-test with a threshold for the t-score of 2.086.

The timeout for the internal evaluation of a single solution was set to 5 minutes (if allowed by the respective tool).
Runs that did not adhere to the given limitations (plus a tolerance threshold) were canceled without considering their results.
That is, the algorithms were canceled if they did not terminate within 110\% of the predefined timeout.
Likewise, the algorithms were killed if they consumed more resources (memory or CPU) than allowed, which happens as overall CPU and memory consumption is hard to control.

In each run, we used 70\% of a randomized, stratified split of the data for learning (search) and 30\% for testing.
We used the \emph{same} splits for all candidates, i.e., for each split and each timeout, we ran each candidate.

The experiments were conducted on (up to) 200 Linux machines in parallel, each of which with a resource limitation of 8 cores (Intel Xeon E5-2670, 2.6Ghz) and 16GB memory.
Thus, the execution of the experimental evaluation took 20.160 CPU hours (840 days) in total.

To ensure a fair comparison especially with respect to hardware resources, all the components required by \tool{}, especially the HTTP servers providing access to the respective libraries, run on the same node.

%\begin{table}[t]
	%\begin{tabular}{l|c|c|c}
	%& WEKA + SK-Learn & WEKA & SK-Learn\\\hline
%    WEKA + SK-Learn & - & 53 & 46 \\\hline
%    WEKA & 86 & - & 60\\\hline
%    SK-Learn & 48 & 35 & -\\
%    \end{tabular}
%    \caption{Preferences for variants of MLS-Plan}
%\end{table}

\subsection{Results}
Table \ref{tbl:results} summarizes the error rates of the different approaches per dataset.
Bold entries indicate that the respective approach achieved the best performance on average within a dataset among the variants of \tool.
To compare \tool in its service variant against the other approaches, we indicate statistically significant improvements of \tool over another approach by $\bullet$ and degradation by $\circ$.

The results show that the performance of the approaches strongly varies across the datasets.
In fact, there is neither a single best approach that is best among most datasets nor is there one approach that is \emph{not} best among at least some datasets.
TPOT seems to dominate on small datasets, but often fails to produce \emph{any} result on larger datasets.

The focus of our evaluation is the comparison of \tool with services from both frameworks WEKA and scikit-learn on one hand and the same composition strategy using algorithms from only one of these libraries on the other hand.
This way, we learn something about the benefit one gains from the service-oriented implementation, which enables combining algorithms from both libraries.
The other three approaches are meant to give reference values of recognized tools in the respective area, but their performance is not relevant for the question whether services are advantageous as their entirely different search behavior is a significant confounding factor.
To isolate the service vs. non-service question from the search strategy, we compare \tool using both frameworks against \tool using only WEKA or scikit-learn respectively.

Note that, a priori, it is completely unclear which of the combinations would be better.
First, applying \tool with both WEKA services and scikit-learn services, does not mean to consider the joint solution space as we consider algorithms occurring in both libraries only once.
That is, many algorithms such as Nearest Neighbors, Random Forests, Naive Bayes, etc. are implemented in both libraries, but in \tool we chose to consider only one of these implementations respectively.
For space limitations, we refer to the documentation of our implementation for the details about the chosen algorithms.
Consequently, the search space of \tool with algorithms of both libraries is not a superset of the search space of the \tool using only WEKA or scikit-learn algorithms respectively.
Second, the search space is still \emph{much larger} compare to using only one of the libraries, which makes it more likely that good solutions need more time to be found. 
While the more powerful search space suggests better solutions due to the coverage of much more pipelines, only a much smaller part of the search space can be examined in the same time bound.

In fact, the above results show that combining the libraries is often significantly better than one or even both of the single-library versions but also worse sometimes.
However, the overall impression is that the service-based variant yields significant improvements, not only over the other \tool-variants but even globally.

We conclude from these results that the automated service composition approach is a relevant approach for solving the \automl problem.
It does not dominate other approaches, but it \emph{is} the best option in quite some cases and should, hence, be in the portfolio of solution techniques.
Therefore, we have demonstrated the utility of automated service composition on the real-world problem of \automl.

\section{Conclusion}
In this paper, we have presented \tool, an approach to automated service composition in the area of machine learning and shown that it can significantly improve the performance in comparison to non-service based approaches.
\tool is based on a reduction of the service composition problem to hierarchical planning with a black box objective function.
The main benefit of using services exploited in this approach is that the service architecture allows to combine algorithms of different frameworks (WEKA and scikit-learn) instead of using only algorithms of one of them, which is the natural limitation one has without the abstraction to the service layer.
The experimental evaluation shows that \tool does bring significant improvements in many cases but also loses against other approaches in some cases, which we trace back to the increased search space size coming with the increased flexibility.
In essence, we interpret our results as an evidence for the utility of automated service composition for the real-world problem of \automl.
However, we also see that the enlarged output space can be a problem, which gives rise to increase timeouts or to improve on the search itself.

%Several topics for future work appear nearby.
%First, the services in our evaluation were hosted on the same machine as the searcher, which brings an artificial and unnecessary limitation of resources that does not exist in this form in the service environment.
%Examining the scaling effect of the composition approach by evaluating much more candidates in parallel would be an important follow-up contribution in the examination of the benefits of service-oriented problem solving.
%Second, on the engineering level, the service managers (GSM) need to be enhanced with respect to data transport.
%Third, for the specific problem of Auto-ML, a long term study with higher timeouts, possibly one week, should be conducted to analyze the benefits of the combined frameworks when more resources are available.
%Finally, we now have provided \emph{one} example domain where service composition shows improvements over non-service approaches, but it would be interesting to identify more such domains.

\bibliographystyle{IEEEtran}
\bibliography{references}

\end{document}